\documentclass[12pt]{article}
\usepackage{amssymb,latexsym,epsf} 
\usepackage{epsf}  
\usepackage{a4wide}

\def\be{\begin{equation}}
\def\ee{\end{equation}}
\def\bea{\begin{eqnarray}}
\def\eea{\end{eqnarray}} 
\def\nn{\nonumber \\}

\def\part{\partial}
\def\tfrac#1#2{{\textstyle{#1\over #2}}}
\def\half{\tfrac{1}{2}}

\def\Z{\ensuremath{\mathbb{Z}}}

\def\cL{{\cal L}}

\def\tg{{\tilde g}}

\def\homega{{\hat{\omega}}}

\def\makeatletter{\catcode`\@=11}
\makeatletter
\def\mathbox#1{\hbox{$\m@th#1$}}%
\def\math@ccstyles#1#2#3#4#5#6#7{{\leavevmode
      \setbox0\mathbox{#6#7}%
      \setbox2\mathbox{#4#5}%
      \dimen@ #3%
      \baselineskip\z@\lineskiplimit#1\lineskip\z@
      \vbox{\ialign{##\crcr
             \hfil \kern #2\box2 \hfil\crcr
             \noalign{\kern\dimen@}%
             \hfil\box0\hfil\crcr}}}}
\def\mathaccstyles{\math@ccstyles\maxdimen}
\def\maththroughstyles{\math@ccstyles{-\maxdimen}}
\def\unitmatrixDT%
 {\maththroughstyles{.45\ht0}\z@\displaystyle {\mathchar"006C}\displaystyle 1}
\begin{document}

\rightline{IFT-UAM/CSIC-00-07}
\rightline{hep-th/0002042}
\rightline{\today}
\vspace{1truecm}

\centerline{\Large \bf Dilatonic Randall-Sundrum Theory} 
\vspace{.3cm}
\centerline {\Large \bf  and renormalization group}
\vspace{1truecm}

\centerline{
    {\bf C\'esar G\'omez,}\footnote{E-mail address: 
                                  {\tt cesar.gomez@uam.es}}
    {\bf Bert Janssen}\footnote{E-mail address: 
                                  {\tt bert.janssen@uam.es}}
    {\bf and} 
    {\bf Pedro Silva}\footnote{E-mail address: 
                                  {\tt psilva@delta.ft.uam.es}}}
\vspace{.4truecm}
\centerline{{\it Instituto de F{\'\i}sica Te{\'o}rica, C-XVI,}}
\centerline{{\it Departamento de F{\'\i}sica Te{\'o}rica, C-XI, }}
\centerline{{\it Universidad Aut{\'o}noma de Madrid}}
\centerline{{\it E-28006 Madrid, Spain}}
\vspace{2truecm}

\centerline{\bf ABSTRACT}
\vspace{.5truecm}

\noindent
We extend Randall-Sundrum dynamics to non-conformal metrics corresponding to
non-constant dilaton. We study the appareance of space-time naked 
singularities and the renormalization group evolution of four-dimensional
Newton constant.

\newpage
\section{Introduction}

The idea that Newtonian gravity can be localized in a three-brane world
has received a lot of attention during the last year \cite{RS1}-\cite{Youm}. 
The original framework is a five-dimensional warped spacetime metric 
of the type:
\be
ds^{2} = A^{2}(z) d{\vec{x}}^{2}  - dz^{2}
\ee
with the warping factor $A$ depending exponentially on $z$, more precisely a
$AdS_{5}$ space-time with negative cosmological constant $\Lambda$. Small 
gravitational fluctuations $h_{\mu\nu}$ of the metric can be written as 
superpositions of modes $h_{\mu\nu}=e^{ipx}\psi(z)\epsilon_{\mu\nu}$.
Four-dimensional gravitons are associated with zero modes defined by the 
condition $p^{2}=0$. In order to get normalizable zero modes
we need to cutoff the deep ultraviolet region of $AdS_{5}$. This can be done
introducing a domain wall at some finite value $z$. On the domain wall metric 
we can now generically have gravitational zero modes that can be interpreted 
as bound states of the higher dimensional graviton strongly localized around 
the wall. This is in summary the dynamical mechanism suggested in 
\cite{RS1, RS2}
to induce four-dimensional Newtonian gravity in a brane world. If we start
with $AdS_{5}$ space-time the resulting domain wall metric will have a horizon
at infinity. Very likely this horizon does  not have any observable effect on 
the physics on the brane due to the strong redshift.

\noindent
From the point of view of holography \cite{Mald}-\cite{Witt}, the 
Randall-Sundrum mechanism of inducing
gravity by introducing an ultraviolet cutoff could be interpreted as the 
extension of the holographic map to conformal field theories
coupled to gravity \cite{BVV, VV}.

\noindent
In this letter we will address the question of extending the RS- scenario to
dilatonic gravity. One reason for that is of course to make a more 
direct contact with string theory where the dilaton appears naturally in the 
definition of brane tensions. Another reason is to unravel how much of 
RS-dynamics depends on conformal invariance. Once we include the dilaton we 
have at our disposal the possibility of working with a vanishing 
five-dimensional cosmological constant. In this case we find bound states 
four-dimensional gravitons with the Newton constant fine tuned in terms of the 
wall tension. For non-vanishing cosmological constant we find a two-parameter
family of solutions depending on the dilaton coupling and on the cosmological
constant. The physics of all these cases is different from that in RS- model
in the sense that in the bulk there appears a naked singularity that can be 
reached from the wall in finite time. This singularity can only be avoided in 
the conformal AdS case. This occurs as an effect of working with a non-constant
dilaton.


\section{Construction of the solutions}

Our starting point is the following five-dimensional action of a dilaton 
$\phi$ coupled to gravity in the presence of a cosmological constant $\Lambda$:
\be
S_{\rm grav}=\frac{1}{\kappa} \int d^4x\ dz \ \sqrt{|G|} \ 
                                    \Bigl[ R -\tfrac{4}{3}(\part \phi)^{2} 
                                        - e^{a \phi/3} \Lambda \Bigr] \ ,
\label{lagran}
\ee 
where $a$ is a free parameter that determines the coupling of the dilaton to 
the cosmological constant. The domain wall solutions of this action are given 
by:
\bea
ds^2 &=& \Bigl[ N(a)\ (z+ z_0) \Bigr]^\tfrac{32}{a^2} d\vec x^2 - dz^2 \ , \nn
e^{-2\phi}&=&\Bigl[ N(a)\ (z+ z_0) \Bigr]^\tfrac{12}{a} ,
\label{solution}
\eea
where $z$ runs between 0 and infinity, $z_0$  is integration constant,
$N(a)$ a function of the coupling $a$ and the cosmological constant:
\be
N(a)= \frac{a^2}{12}\ \sqrt{\frac{3\Lambda}{(a^2-64)}} \ .
\label{N}
\ee
Since we assume the cosmological constant to be negative, our solution only 
makes sense for $a$'s with values between 8 and $-8$.  
The above metric has clearly a naked singularity at 
$z=-z_0$:
\bea
&&R = 2^7\ \frac{40 - a^2}{a^4 \ (z+z_0)^2}      \  ; \\
&&R_{\mu\nu\rho\lambda}R^{\mu\nu\rho\lambda} =
   2^{13}\ \frac{640-32a^2+a^4}{a^8 \ (z+z_0)^4}\ .
\eea
To calculate the profile of the graviton we add some small fluctuations 
$h_{\mu\nu}$ to the above background metric, choosing the
gauge $h_\mu^\mu = \part^\mu h_{\mu\nu}=h_{z\nu}=h_{zz}=0$:
\be
ds^2 = \Bigl[ N(a)\ (z+ z_0) \Bigr]^\tfrac{32}{a^2}
                 (\eta_{mn}+h_{mn})\ d x^m dx^n - dz^2 \ ,
\label{perturbed}
\ee
Inserting the perturbed metric (\ref{perturbed}) in the equations of motion of
(\ref{lagran}), we get, at first order in the perturbation, the following 
differential equation for the graviton:
\be
\Bigl[\ \half \part^2 
     +\frac{16}{a^2}\cdot\frac{32-a^2}{a^2}\  (z+z_0)^{-2}\ \Bigr] h_{mn}=0 
\ee
For the splitting of variables $h_{\mu\nu}(x,z)=\ e^{i\vec k\vec x}\ \psi(z)\
\epsilon_{\mu\nu}$, we find for the profile of the graviton zero mode
\be
\psi(z) = \Bigl[ N(a)(z+z_0) \Bigr]^{32/ a^2}. 
\label{zeromode}
\ee
This zero mode is not normalizable over the whole of space. If we insist on the
existence of a graviton zero mode, we need to introduce a cut-off at the 
position $z=N(a)^{-1}$.

\noindent
The cutt-off has the effect of throwing away the part of space with 
$z>N(a)^{-1}$, where the graviton zero mode (\ref{zeromode}) becomes 
non-normalizable.
We can replace this thrown away part by a copy of the part of space with 
$z<N(a)^{-1}$. At the level of the solution (\ref{solution}), this is seen in 
the fact that we pass from the variables  $z \rightarrow N(a)^{-1}- |z|$, 
(where $|z|$ now runs between 0 and $N(a)^{-1}-z_0$), where
$N(a)z_0 \in [0, 1]$.   This generates 
a delta function behaviour in the equations of motion, which can be compensated
by introducing domain wall source terms at the bounderies:
\bea
S_{\rm source} &=& \int_{z=0} d^4x \sqrt{|\tg|}\ 
                   \Bigl[ \cL_{\rm brane} + e^{b\phi/3}V_0 \Bigr] \nn
          &&  \hspace{1cm}
              +    \int_{z=N(a)^{-1}} d^4x \sqrt{|\tg|}\ 
                   \Bigl[ \cL_{\rm brane} + e^{b\phi/3}V_L \Bigr]\ ,
\label{source}
\eea
where $\tg_{mn}=G_{\mu\nu}\delta^\mu_m\delta^\nu_n$ is the induced metric on 
the domain wall, 
$\cL_{\rm brane}$ is the Lagrangian of a gauge theory living an the 
brane and $V_i$ the tensions of the branes. We thus get for the solution of 
the space with domain wall:
\bea
ds^2 &=& \Bigl[ 1- N(a)\ |z| \Bigr]^\tfrac{32}{a^2} d\vec x^2 - dz^2 \ , 
                         \nn
e^{-2\phi}&=&\Bigl[1- N(a)\ |z| \Bigr]^\tfrac{12}{a} ,
\label{|solution|}
\eea
The brane tensions $V_i$ and the dilaton coupling $b$ satisfy the matching 
conditions
\be
V_0=-V_L = \frac{8}{\kappa} \sqrt{\frac{3\Lambda}{a^2-64}}\ ,
\hspace{2cm}
b=\half a \ .
\ee
$V_0$ ($V_L$) corresponds to the tension of the so-called Planck brane 
(TeV-brane). What we are actually doing by introducing these source terms is 
making an orbifold construction $S^1/\Z_2$, where the domain walls are 
located at the fix points. Note that we can also take the limit in which we 
send the TeV-brane to the singularity, by taking the limit $z_0 \rightarrow 0$.

\noindent
In these variables the Randall-Sundrum (RS) limit $a \rightarrow 0$ 
is singular. To get a good picture of this limit, it is instructive to go to 
the conformal frame via the coordinate transformation
\be
\Bigl[ 1- N(a) |z| \Bigr]^{\tfrac{a^2-16}{a^2}} = 
            \Bigl[1 - O(a) |\omega| \Bigr]  
\label{coordtransf}
\ee 
where $|\omega|$ runs between 0 and 
$\omega_0= O(a)^{-1} \Bigl[1-\Bigl(N(a) z_0\Bigr)^\tfrac{a^2-16}{a^2} \Bigr]$ 
and $O(a) = \frac{a^2-16}{a^2} N(a)$. In this frame the solution takes  the
form 
\bea
ds^2&=& \Bigl[1- O(a)|\omega| \Bigr]^\tfrac{32}{a^2-16}
                      (d\vec x^2 - d\omega^2) \ , \nn
e^{-2\phi}&=&\Bigl[1- O(a)|\omega| \Bigr]^\tfrac{12a}{a^2-16} .
\label{conformalsol}
\eea


\begin{figure}
\begin{center}
\leavevmode
\epsfxsize=12cm
\epsffile{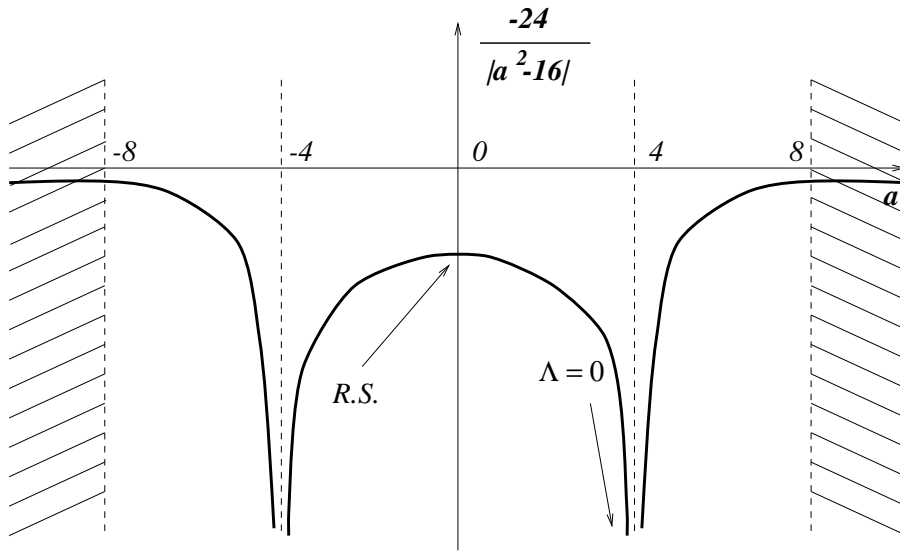}
\caption{\it The graphic of the power of the graviton zero mode 
(\ref{graviton2})-(\ref{graviton3}). There exists always a coordinate frame in 
which the zero mode can be described as being confined. In the region $-4<a<4$ 
this frame corresponds to the conformal one. The point $a=0$ corrsponds to the 
Randall-Sundrum solution. The solution for $a=4$ only exists for $\Lambda=0$. }
\label{gravit}
\end{center}  
\end{figure}
\begin{figure}
\begin{center}
\leavevmode
\epsfxsize=12cm
\epsffile{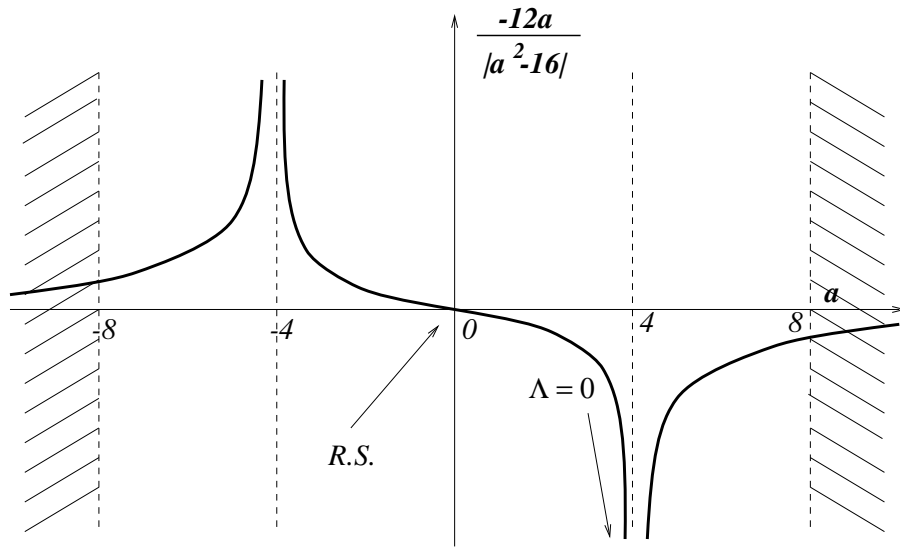}
\caption{\it The graphic of the power of the dilaton (\ref{|solution|})
-(\ref{dilaton3}). Again the region $-4<a<4$ corresponds to the conformal 
frame. For positive values of $a$ the dilaton is proportional to the graviton,
while for negative $a$'s the dilaton is inversely proportional. }
\label{dilat}
\end{center}  
\end{figure}

We can now make a case study for the different values of the dilaton couping 
$a$: 
\begin{itemize}
\item In the conformal frame the RS limit $a\rightarrow 0$ is prefectly 
regular and gives us the non-dilatonic AdS${}_5$ solution of RS 
\cite{RS1, RS2}. The graviton zero mode goes like 
$\psi(\omega)=\Bigl[ 1-\sqrt{\tfrac{\Lambda}{12}}\ |\omega| \Bigr]^{-3/2}$\ .

\item For $0<a^2<16$ we find the graviton zero mode falling off like
\be
\psi(\omega)= \Bigl[ 1- O(a)|\omega| \Bigr]^\tfrac{24}{a^2-16}\ ,
\label{graviton2}
\ee 
i.e. confining faster and faster as $a$ approaches the value $-4$. Note 
that the RS limit is the least confining case of this family.

\item For $16<a^2<64$ we find that the zero mode is normalizable only in a 
finite interval, even in the limit $\omega_0 \rightarrow O(a)^{-1}$
\footnote{Note that by the definition of the range of $\omega$,  
$\omega_0 = O(a)^{-1}$ is the maximal value it can attain.}. However,
we can always change coordinates and make the interval $[0, \omega_0]$ 
infinite. For example in coordinates $1-O(a)\homega = (1-O(a)\omega)^{-1}$.
Again the zero mode is confining
\be
\psi(\homega)= \Bigl[ 1- O(a)|\homega| \Bigr]^\tfrac{-24}{a^2-16}\ ,
\label{graviton3}
\ee
and the dilaton in these coordinates looks like
\be
e^{-2\phi}=\Bigl[1- O(a)|\homega| \Bigr]^\tfrac{-12a}{a^2-16} .
\label{dilaton3}
\ee
Note that the exponent of the graviton zero mode is bigger (smaller) than in 
the RS case for values of $a<\sqrt{32}$ ($a>\sqrt{32}$). However, a comparison 
as in the previous case is difficult since after the coordinate transformation,
we are no longer in the conformal frame.
  
\end{itemize}

\noindent
Concluding, we find that in the any of the cases discussed above there is 
confinement of the gravitational zero mode, in some cases even stronger than 
in the RS-case. However there is a big difference in the behaviour of the 
dilaton: for positive values of $a$ the exponent of the dilaton has the same 
sign as the exponent of the graviton, but for negative value of $a$ the signs 
are opposite.
Depending on the sign of $a$ the string coupling constant $e^{\phi}$ at the
space-time singularity goes to $\infty$ or zero.

\noindent
Although the RS limir $a\rightarrow 0$ is well defined in the conformal frame
(\ref{conformalsol}), there is a singular point for the value $a=\pm4$, which 
needs a special analysis. Solving the equations of motion for the $a=4$
case, it becomes clear that there are only two solutions:
either $\Lambda = 0$ or linear dilaton with constant warp factor i.e flat 
five-dimensional space-time metric. If we consider non-critical
strings in five dimensions the cosmological constant term is given, in string
frame, by
$e^{-2\phi}\frac{(D_{cr}-D)}{3}$ with $D_{cr} = 26$ or 10. In this case the 
only solution is flat five-dimensional space-time and dilaton:
\bea
\phi = \frac{1}{2}\sqrt{\Lambda}z\ .
\eea
For $\Lambda = 0$ we 
find two solutions for two distict values of the coupling $b$, which 
only differ in the dilaton dependence: 
\bea
&b=-4:& \ \left\{ \begin{array}{ll}
                 ds^2 =\Bigl[1- \tfrac{2}{3}\kappa V_0 |z| \Bigr]^\tfrac{1}{2} 
                             d\vec x^2 - dz^2 \ , \\
                         \\
                 e^{-2\phi}=
              \Bigl[1- \tfrac{2}{3} \kappa V_0|z| \Bigr]^{-\tfrac{3}{2}} ,
               \end{array}
         \right. 
\label{Lambda0-1}\\ 
\nn \nn
&b=4:& \ \left\{ \begin{array}{ll}
                ds^2 =\Bigl[1- \tfrac{2}{3} \kappa V_0 |z| \Bigr]^\tfrac{1}{2} 
                     d\vec x^2 - dz^2 \ , \\
                         \\
                e^{-2\phi}=
                   \Bigl[1- \tfrac{2}{3} \kappa V_0|z| \Bigr]^\tfrac{3}{2} .
               \end{array}
          \right.
\label{Lambda0-2}
\eea 
Here the coordinate $z$ runs between 0 and 
$\Bigl( (\tfrac{2}{3}\kappa V_0)^{-1}  -z_0\Bigr)$ where 
$z_0 \in [0, (\tfrac{2}{3}\kappa V_0)^{-1} ].$ In the conformal frame, these 
solutions are of the form:  
\bea
&b=-4:& \ \left\{\begin{array}{ll}
                ds^2 =\Bigl[1- \half \kappa V_0 |\omega| \Bigr]^\tfrac{2}{3} 
                     (d\vec x^2 - d\omega^2) \ , \\
                         \\
                e^{-2\phi}=\Bigl[1- \half \kappa V_0|\omega| \Bigr]^{-2} ,
               \end{array}
       \right. 
\label{confLambda0-1}\\ 
\nn \nn
&b=4& : \ \left\{\begin{array}{ll}
                ds^2 =\Bigl[1- \half \kappa V_0 |\omega| \Bigr]^\tfrac{2}{3} 
                     (d\vec x^2 - d\omega^2) \ , \\
                         \\
                e^{-2\phi}=\Bigl[1- \half \kappa V_0|\omega| \Bigr]^{2} .
               \end{array}
       \right.
\label{confLambda0-2}
\eea
Again, as in the case of $16<a^2<64$ above, we can always find a coordinate 
system in which the graviton zero mode is confined:
\be
\psi(\homega)= \Bigl[1- \half \kappa V_0 |\homega| \Bigr]^{-\tfrac{1}{2}} \ .
\ee 
The other singular point is at $a=-4$. This singularity however turns out to 
be a coordinate singularity due to the singular behaviour of the coordinate
transformation (\ref{coordtransf}). The solution at this point is given by
\bea
ds^2&=& e^{-\sqrt{-3/16\Lambda}|\omega|} \Bigr(d\vec x^2 - d\omega\Bigl) \ ,
\nn
e^{-2\phi}&=& e^{-\tfrac{3}{2}\sqrt{-3/16\Lambda}|\omega|}\ .
\eea
Note that also in this case the graviton zero modes are confined.

\section{Renormalisation group and Newton constant}

\noindent
Recently a different approach to RS-dynamics based on renormalization group
interpretation of holography \cite{Akh, AG} has been suggested in reference 
\cite{BVV, VV}. In this approach the Einstein gravity on the wall is replaced 
by the
integral on the ultraviolet
region of the five-dimensional effective action. The four-dimensional
cosmological constant
remains fixed along the renormalization group evolution and therefore can be
fine tuned to zero by impossing appropiated boundary conditions in the
ultraviolet region. The solutions we
have been describing above correspond to particular initial conditions 
determined by the values
of the wall tension. Notice that this value fixed by the jump equations is
independent
of the particular value of the UV cutoff used to locate the wall i.e. it is
renormalization
group invariant.

\noindent
In this renormalization group scheme we can define the following beta
function:
\be
\beta_{\phi} = A\frac{\partial}{\partial A} \phi\ ,
\ee
or in terms of the ``cosmological time'':
\be
\gamma = \frac{\dot{\phi}}{\beta_{\phi}} \ ,
\ee
with
\be 
\gamma = \frac{\dot{A}}{A} 
\ee
the expansion rate of the four-dimensional metric.
For the solutions of dilatonic gravity we have:
\be
\beta_{\phi} = -\frac{3}{8}a \ ,
\ee
and we observe that the RS-model  $a =0 $ corresponds to the
conformal case
$\beta_{\phi} =0$ with all other cases constant but non-vanishing beta
function (positive or negative
depending on the sign of $a$). The naked singularity is 
characterized by infinite $\gamma$ i.e by $\dot{\phi} = \infty$. For the
vanishing cosmological 
constant case we get the beta function:
\be
\beta_{\phi} = \frac{3}{2}\ ,
\ee
corresponding precisely to the point $a=4$ i.e to the singular
line in Figure \ref{gravit} and \ref{dilat}.

\par

Next we will study the evolution of the Newton constant. The relevant
renormalization group equations is given by:
\be
(\dot{A}\frac{\partial}{\partial A} +
\dot{\phi}\frac{\partial}{\partial{\phi}}) \frac{1}{\kappa_4} =
\frac{A^{2}}{\kappa_{0}}\ .
\ee
This equation have a very simple physical meaning. Namely the r.h.s of the
equation is simply the ``time'' derivative of the Newton constant $\kappa_{4}$ 
defined by Kaluza-Klein reduction on the bulk direction. Thus
the meaning of the previous equation is simply that $\frac{\partial
\kappa_{4}}{\partial t} =0$.

\noindent
This equation, once we have 
written $\dot{\phi}$ and $\dot{A}$ in terms of $A$ has the the solution,
\be
\kappa_4^{-1}= \frac{1}{\kappa} \int dz \ \ A^2(z) + {\rm constant}\ .
\ee 
For $A$ as in (\ref{|solution|}), the above integral becomes
\be
\kappa_4^{-1}= \frac{1}{\kappa} \int_0^{N(a)^{-1}-z_0} dz 
                           \Big[ 1- N(a)|z| \Bigr]^\tfrac{32}{a^2}\ .
\ee 
In order to be able to compare with the case of zero-dilaton (Randal-Sundrum),
we switch to the conformal variable $\omega$,  
\be
\kappa_4^{-1}= \frac{1}{\kappa} \int_0^{\omega_a} d\omega  
                           \Big[ 1- O(a)|\omega| \Bigr]^\tfrac{32}{a^2} \ ,
\label{integral} 
\ee
where the upper limit corresponds to the distance between the Planck brane and
the place where the effective Newton constant is measured.  For the case 
$a^2<16$, $\omega_0$ runs in a semi-infinite range. Solving the integral
(\ref{integral}) gives
\be
\kappa_4^{-1}= \frac{1}{\sqrt{-3\kappa^2\Lambda}}\ 
              \frac{12\sqrt{64-a^2}}{32+a^2} \  
         \Bigl[ 1- \Bigl(1-O(a)\omega_0\Bigr)^{-\tfrac{32+a^2}{16-a^2}} \Bigr] 
\ ,
\ee
which should be compared to the RS case:
\be
\kappa^{-1}_{4\ (RS)}= \sqrt\frac{-3}{\kappa^2\Lambda} 
         \Bigl[ 1 - \Bigl(1+\sqrt{\frac{-\Lambda}{12}}\ \omega_0\Bigr)^{-2} 
\Bigr] \ .
\ee
On the other hand, it is not straightforward how to give an adequate
comparison for case $16<a^2<64$ and Randall-Sundrum. We have mentioned above 
that in the conformal 
frame the variable $\omega$ then runs over a finite range. This range can be 
made infinite, as done above, but again comparison to RS is hard due to the 
fact that we are no longer in the conformal frame.  

\begin{figure}
\begin{center}
\leavevmode
\epsfxsize=10cm
\epsffile{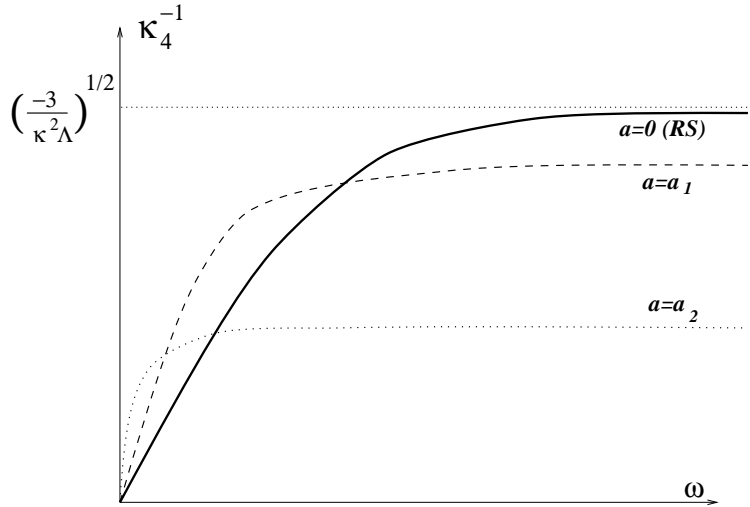}
\caption{\it The running of the effective four-dimensional Newton constant as 
a function of the holographic distance $\omega$ for different values of 
the dilaton coupling $0<a_1^2<a_2^2<16$.}
\label{kappa}
\end{center}  
\end{figure}

\noindent
The other physical implication is the screening of the measurements of 
physical quantities at the distance $z_0$ by the warp factor $A(z)$, generating
a hierarchy between the Planck brane and the brane TeV-brane. In terms of the 
four and five-dimensional Planck-length the hierarchy is of the order of
\be
\ell_4 \approx e^{-50} \ell_{5} \ .
\ee
In the conformal frame (the only frame where we can compare to the RS case), 
we get
\be
\omega_0= -\tfrac{12}{a^2-16} \sqrt{\tfrac{a^2-64}{3\Lambda}}
\Big(e^{\frac{25}{8}(16-a^2)}-1\Big)\ ,
\ee  
which should be compared to the Randal-Sundrum case,
\be
\omega_0=\sqrt{\frac{-12}{\Lambda}}\Big(e^{50}-1\Big)\ .
\ee
We see that the inclusion of the dilaton has a considerable effect on the 
effective Newton constant: the higher the values of dilaton coupling $a^2$, the
faster the Newton constant reaches its asymptatic value. At the same 
time there is a screening of the constant which is bigger as the dilaton
coupling grows.

Finally let us just mention a natural interpretation of
the singularity from the four-dimensional physics point of view. This 
singularity, depending on the sign of $\beta_{\phi}$ could be interpreted
either as a Landau pole or a confinement scale for the non-conformal gauge 
theory on the wall. It is interesting to see that this potential scale of
the gauge theory is related with the four dimensional Newton scale.

\vspace{1cm}

\noindent
{\it NOTE ADDED IN PROOF:}
While this paper was being written we received the papers \cite{rest, KSS}
that partially overlap with our results. 

\vspace{1cm}
\noindent
{\bf Acknowledgments}\\
The work of C.G. and B.J. has been supported by the TMR program FMRX-CT96-0012 
on {\sl Integrability, non-perturbative effects, and symmetry in quantum field 
theory}. The work of P.S. was partially supported by the gouverment of 
Venezuela.


\end{document}